\documentclass[%
reprint,
superscriptaddress,
showpacs,
 amsmath,amssymb,
 aps,
prb,
twocolumn,
oneside,
floatfix,
]{revtex4-1}

\usepackage{graphicx}
\usepackage{dcolumn}
\usepackage{bm}
\usepackage[utf8]{inputenc}
\usepackage{amssymb,bbm}
   
\usepackage{braket}
\usepackage{xcolor}
\usepackage{color}
\usepackage{soul}

\newcommand{\si}{\sigma}
\newcommand{\eps}{\epsilon}
\newcommand{\+}{^{\dagger}}

\newcommand{\nh}{\hat{n}}
\newcommand{\ua}{\uparrow}
\newcommand{\da}{\downarrow}
\newcommand{\blue}[1]{#1}

\begin{document}

\preprint{APS/123-QED}

\title{Nature of the Doping-Driven Metal-Insulator Transition in Correlated Electron Systems with strong Hund's Exchange Coupling}

\author{Jakob Steinbauer}
 \email{jakob.steinbauer@polytechnique.edu}
  \affiliation{CPHT, CNRS, Ecole Polytechnique, Institut Polytechnique de Paris, F-91128 Palaiseau, France}
 \author{Luca de' Medici}%
 \email{luca.demedici@espci.fr}
\affiliation{%
\'Ecole Supérieure de Physique et Chimie Industrielles de la Ville de Paris
}%
\author{Silke Biermann}%
 \email{silke.biermann@polytechnique.edu}
 \affiliation{CPHT, CNRS, Ecole Polytechnique, Institut Polytechnique de Paris, F-91128 Palaiseau, France}
 \affiliation{Coll\`{e}ge de France, 11 place Marcelin Berthelot, 75005 Paris, France}
 \affiliation{European Theoretical Spectroscopy Facility, Europe}

\date{\today}

\begin{abstract}
We study the doping-driven Mott metal-insulator transition for multi-orbital 
Hubbard models with Hund’s exchange coupling at finite temperatures.
As in the single-orbital Hubbard model, the transition is of first-order 
within dynamical mean field theory, with a coexistence region where two
solutions can be stabilized. \blue{We find, that in the presence of finite Hund's coupling, the insulating phase is connected to a badly metallic phase, which extends to surprisingly large dopings.}
While fractional power-law behavior of the self-energies on the Matsubara axis 
is found on both sides of the transition, a regime with frozen local moments 
develops only on the branch connected to the insulating phase.
  
\end{abstract}

\pacs{Valid PACS appear here}
\maketitle


\section{\label{sec:Introduction} Introduction}

Correlated materials feature some of the most fascinating phenomena of condensed matter physics, such as exotic transport phenomena including metal-insulator transitions\cite{Mott} or superconductivity\cite{strong_corr_SC, Cuprates_SC,pnictide_SC}, satellite features\cite{werner_nature}  in spectroscopic measurements, or spin-, charge-\cite{kennedy_lieb_order}, or orbital ordering\cite{Orbital_ordering_Long}.
The single-orbital Hubbard model offers a minimal description \blue{of many of} the effects associated with strong correlations. \blue{ Among its most prominent results is the prediction of localisation of the conduction electrons at half filling when the Coulomb repulsion, parametrized by the Hubbard interaction $U$, dominates over the kinetic energy of the electrons, thus causing a metal-insulator transition\cite{BrinkmanRice,MIT_DMFT_RozenbergKotliar}.} In the proximity of this transition, unconventional -- correlation-driven -- phenomena are found, reminiscent of some of the exotic properties observed in transition metal oxides \cite{imada}. \par
For most materials, however, the single band description of the Hubbard model is a crude over-simplification, and an even richer physics results from the additional degrees of freedom present in multi-orbital models.
Characteristic changes in orbital occupations drive e.g.
the metal-insulator transitions in vanadium sesquioxide
\cite{PoteryaevPRB2007} 
or vanadium dioxide \cite{BiermannPRL2005}. 
Indeed, the {\it effective degeneracy} of a material, that is, the number
of active orbitals {\it after} correlations have been included,
has been identified as a determining factor for its properties
\cite{pavarini:176403, 
MartinsPRL2011, 
MartinsJPCM2016}. 
But even in metallic phases, orbital physics can play a role: 
The properties of iron pnictides and chalcogenides, or 4d transition metal oxides such as ruthenates, for instance, can only be described accurately when considering their multi-orbital nature. Here, a key parameter is the reduction of the on-site effective Coulomb interaction for electrons with parallel spins due to Hund's exchange coupling. This same quantity parametrizes -- for antiparallel spins -- the difference between the effective on-site intra-orbital and inter-orbital interactions. 
For large Hund's coupling $J$, the conduction electrons appear to be strongly
correlated - with large quasi-particle masses and low coherence
scales - without being in the immediate vicinity of the Mott transition,  \blue{ i.e. the Hubbard interaction $U$ only being a moderate
fraction of the critical value $U_c$ at which the Mott transition takes place.}
For the past few years, such ``Hund's metals'' have been the subject of intensive studies, both on a model level\cite{Fanfarillo_Bascones, Pruschke2005, STADLER2018, janus_faced_prl, spinfreezing_troyermillis, demedici_Hunds_metals, Hundreview} and within realistic calculations for specific materials \cite{Haule_Kotliar_NewJPhys, Yin_Haule_Kotliar_Nature, demedici_Hunds_metals, Hundreview, werner_nature}.
The main findings are \blue{a change of the critical interaction $U_{c}$, marking the Mott metal-insulator transition; decreasing at half filling while increasing at other commensurate fillings}
, a general decrease of coherence, as well as a tendency of orbital decoupling with increasing $J$. \par
Another interesting feature, first discussed in a work by Werner et
al.\cite{spinfreezing_troyermillis}, is the occurrence of
an extended region in the parameter space spanned by filling $n$,
temperature $T$ and interaction $U$, where spin-spin correlation functions do not decay
and self-energies display fractional power-law behavior or extrapolate to significant finite values on the Fermi surface.
\blue{Early on\cite{spinfreezing_troyermillis}, this fractional power-law behavior of the self-energies thought to be related to the peculiar optical conductivity of various ruthenates\cite{SrRuO3,SrRuO3_2,SrRuO3_3}. In Ref. \onlinecite{werner_nature}, such a behavior -- reminiscent of non-Fermi liquid phases in orbital-selective Mott
insulators \cite{Biermann_Medici_nonfermi} -- was even evidenced in realistic materials simulations for
the iron pnictide BaFe$_2$As$_2$. }\par
\blue{Most interestingly, recent works\cite{Hund_stadler,Jernej_Kondo_Kanamori, Jernej_SOC, STADLER2018} suggest that these findings are just an aspect of a more general phenomenon -- spin-orbit separation -- which is not restricted to lattice models, but already emerges from the physics of the multi-orbital impurity model with Hund's coupling. In the multi-orbital Anderson impurity model, the competition between Hund's coupling and Kondo screening gives rise to an intermediate regime, where the Kondo scales of orbitals $T_{K}^{orb}$ and spins $T_{K}^{sp}$ are distinct\cite{Jernej_Kondo_Kanamori}. In the intermediate energy range $T_{K}^{orb} > |\omega| , T > T_{K}^{sp}$, orbital degrees of freedom are screened by the conduction electrons. This gives rise to a large local spin moment, at which conduction electrons are scattered, and results in strong decoherence and non-Fermi liquid behavior. Only below $T_{K}^{sp}$, Fermi liquid behavior is recovered\cite{deLeoPhD, Hundreview,  Hund_stadler, STADLER2018, Sangiovanni}. In Ref. \onlinecite{Hund_stadler} this phenomenon was investigated for a three-orbital impurity model with a flat density of bath states, as well as within dynamical mean field theory on the Bethe lattice. In both cases, the described physics was found for fillings $1 < n < 2N_{orb}-1$, with the spin-orbit separation becoming stronger upon approaching half filling.}\par
The \blue{ spin-freezing phenomenon, as well as orbital-freezing for negative Hund's $J$\cite{karimsteiner_longrange},} is of particular interest due to its possible connection with the appearance of superconductivity; the enhancement of local magnetic fluctuations characterizing the spin-freezing boundary has been argued to result in an effective attraction between electrons, even for repulsive bare interactions\cite{Hoshino_Werner_SC_magn_mom,werner_spin_cuprates}. \par
In addition, 
a recent model study\cite{Hundsinstab_medici} using the slave-spin technique\cite{Slavespin_medici} found Hund's coupling to cause an instability of the Fermi-liquid, preceded by a divergence of the electronic compressibility. It was argued that this might have  important consequences, like a tendency of the system towards phase separation or a strong enhancement of the quasi-particle interaction vertices in the homogeneous phase. \par
\blue{ Despite these recent advances in the understanding of the consequences of Hund's exchange,
it remains unclear how this connects with and influences the existence of the finite-temperature metal-insulator coexistence region\cite{Doping_1orb_wernermillis}.} 
In this work, we tackle some of these questions, by investigating the finite doping properties of multi-orbital models with Hund's coupling around half filling. These models display a first-order Mott metal-insulator transition which is characterized by an extended coexistence region where both metallic and insulating phases can be realized. We study the nature of these phases, focusing on
the electronic compressibility and susceptibility \blue{ at} finite temperatures, and analyzing the evolution of spin-freezing phenomena as a function of doping and temperature.

The paper is structured as follows: Section \ref{sec:model} describes the model chosen for the present study. Section III presents our results, with emphasis on the phase diagram (subsection III.A), temperature-dependence (subsection III.B), spin-spin correlations (subsection III.C) and the role of spin-flip and pair-hopping
terms in the interaction part of the Hamiltonian (subsection III.D).
In Section IV we present a summary and conclusions.

\section{\label{sec:model}The Model}
\subsection{Model and methodology}
The model under consideration is a multi-orbital Hubbard Hamiltonian with Hund's coupling $\hat{H} = \hat{H}_{0} + \hat{H}_{int}$, with 
\begin{align}\label{eq:H0}
  \hat{H}_{0}&=t\!\sum_{\Braket{ij}m\si} 
 c\+_{im\si}c_{jm\si}-(\mu+\mu_{0})\sum_{im\si}\nh_{im\si}
\end{align}
where $c\+_{im\si}$ ($c_{im\si}$) creates (annihilates) an electron of orbital $m$ and spin $\si$ at site $i$. \blue{ The chemical potential $\mu$ is shifted by a term $\mu_{0}$, such that $\mu=0$
corresponds to a half-filled system.}
We assume orbital-diagonal hopping amplitudes $t$.
\blue{Considering} the lattice geometry, for reasons of simplicity, we choose a Bethe
lattice with an infinite coordination  number, corresponding 
to a semi-elliptic density of states
\begin{align}
\mathcal{D}(\omega) = \theta(2t-|\omega|)\frac{\sqrt{4t^{2}-\omega^{2}}}{2\pi t^{2}} \text{ .}
\end{align}
The interaction term is
\begin{align}\label{eq:Hint}
\begin{split}
\hat{H}_{int}&=U\sum_{m} \nh_{m,\ua}\nh_{m,\da}\\
&+U'\!\sum_{m > m' \si} \nh_{m\si}\nh_{m'\bar{\si}} +(U'-J)\!\sum_{m > m' \si}\nh_{m\si}\nh_{m'\si}\\
& + \alpha\! \left[
  J\!\sum_{m\neq m'}\! c\+_{m\ua}c\+_{m\da}c_{m'\da}c_{m'\ua}
  -J\!\sum_{m\neq m'}\!c\+_{m\ua} c_{m\da} c\+_{m'\da}c_{m'\ua}  
  \right] \text{ .}
\end{split}
\end{align}
\blue{For reasons of computational simplicity, 
we drop the spin-flip and pair-hopping terms, i.e. we set $\alpha=0$, for all calculations, except those presented in section \ref{non-nn}.}
For the remaining density-density terms of the interaction, we choose the standard form
 \cite{kanamori} $U' = U-2J$, with a positive Hund's exchange coupling $J>0$.
We assume that all terms in \eqref{eq:Hint} contribute non-negative repulsive interactions, restricting us to $J<U/3$. \blue{ Throughout this work, we will consider a typical Hund's coupling of $J=0.25U$.}\\
We solve the model using dynamical mean field theory (DMFT), which maps the model onto a single site,
coupled to a self-consistent bath \cite{DMFTkotliar_georges}. \par
The resulting quantum impurity model is solved using the continuous time hybridization-expansion (CT-Hyb) quantum Monte Carlo algorithm\blue{; for a detailed discussion of various quantum Monte Carlo algorithms consider Ref. \onlinecite{ctQMC}.} In general, solutions were obtained from $\sim 10^{6}$ Monte Carlo sweeps per core, \blue{using 40 cores} and $50-100$ DMFT iterations. Close to the spinoidals, the number of sweeps was increased to  $\sim 4\times 10^{6}$, with hundreds of DMFT iterations at a mixing factor of $0.4$. \par
\blue{ Since we are interested in the coexistence region, we have to stabilize two different solutions for the same set of parameters. In the framework of DMFT, this is achieved by using the converged hybridization functions from previous simulations as starting points for the new Monte Carlo calculations. In the course of this paper, we will use the term ``metallic'' for solutions stabilized using this scheme by starting in the strongly doped regime (large $\mu$) upon successively decreasing the chemical potential, while solutions obtained upon increasing the chemical potential away from half filling will be denoted as ``insulating'', irrespective of their specific characteristics (e.g. spectral weight at the Fermi level).}\par
From now on, we will work with dimensionless variables, expressing all energies
in units of the hopping $t$, which we set to $t=1$.

\subsection{The multi-orbital Hubbard model: A short reminder}
The multi-orbital Hubbard model has been the subject of numerous previous
studies\cite{Hund_stadler, Pruschke2005, spinfreezing_troyermillis, janus_faced_prl,vDelft_multiorbital, Multi_orbital_von_Delft, two_orbital_mott_transition_kawakami, rozenberg_multiorb}. In the limit of infinite coordination number considered here, DMFT
yields the exact solution, unless specific approximations are introduced by
the choice of an approximate solver technique. At commensurate fillings and
low enough temperature,
the system undergoes a metal-insulator transition as a function of the interaction
strength \cite{DMFTkotliar_georges}.
  Metallic or insulating solutions are characterized respectively by a finite or vanishing spectral weight at the Fermi energy.
  In the one-band Hubbard model\cite{DMFTkotliar_georges}, as well as in the multi-band Hubbard models in the absence of Hund's coupling\cite{rozenberg_multiorb}, it is known that at half filling and at low enough temperatures a metallic solution exists for $U<U_{c2}(T)$, whereas a Mott insulating solution exists for  $U>U_{c1}(T)$. Since it is found that $U_{c1}(T)<U_{c2}(T)$, there is a parameter zone where the two solutions coexist. The Mott transition is thus a first-order transition, taking place inside this zone at the value of $U_c(T)$ where the free energies of the two solutions are equal. This first-order line terminates in two second-order critical points, one at zero temperature, where $U_{c}(T=0)=U_{c2}(T=0)$, and the other one at a finite temperature $T_c$ where the two lines delimiting the coexistence region meet, i.e. $U_{c1}(T_c)=U_{c}(T_c)=U_{c2}(T_c)$.
  This scenario is by now established and provides a model description for the interaction-driven Mott transition. \par
  An alternative path in parameter space allows one to study doping-driven transitions, varying the chemical potential away
  from the electron-hole symmetric value.
  We focus on the evolution of the metallic and insulating phases coexisting in the doping-driven transition.\par
  The physics in this region depends on the branch -- metallic or insulating -- considered.
  On the metallic side, a variation of the chemical potential is immediately accompanied by a corresponding change in the
  filling, expressing the finite compressibility of the system.
  At zero temperature, the insulating phase remains half filled as long as the chemical potential stays inside the gap.
  Upon further increase of $\mu$, this solution eventually ``jumps'' to coincide with the metallic one,
  and the insulating branch disappears.\par
   At finite temperatures, the coexistence region encompasses a volume
  in the parameter space $(U,T,\mu)$\cite{compressdiv_kotliar}. The insulating branch then accommodates
  tiny values of doping, thus realizing a strongly incoherent metal. 
  For values of $U>U_{c2}$ and a chemical potential sufficiently close to the electron-hole symmetric value, only an insulating
  solution exists. It is delimited by two coexistence regions
  corresponding to the first-order Mott transitions under electron- and hole-doping respectively.
  \cite{DMFTkotliar_georges,compressdiv_kotliar}.
  The general aspects of this scenario survive in the context of multi-orbital systems at least as long as only
  SU($N$)-symmetric\cite{vDelft_multiorbital} interactions are considered. The fate of the respective
  phases in the presence of Hund's exchange coupling, however, remains an open question.
  In this work, we investigate
  the physics of the coexistence region in the presence of a finite Hund's coupling.
\begin{figure}[t]
\begin{center}
\includegraphics[scale=0.24]{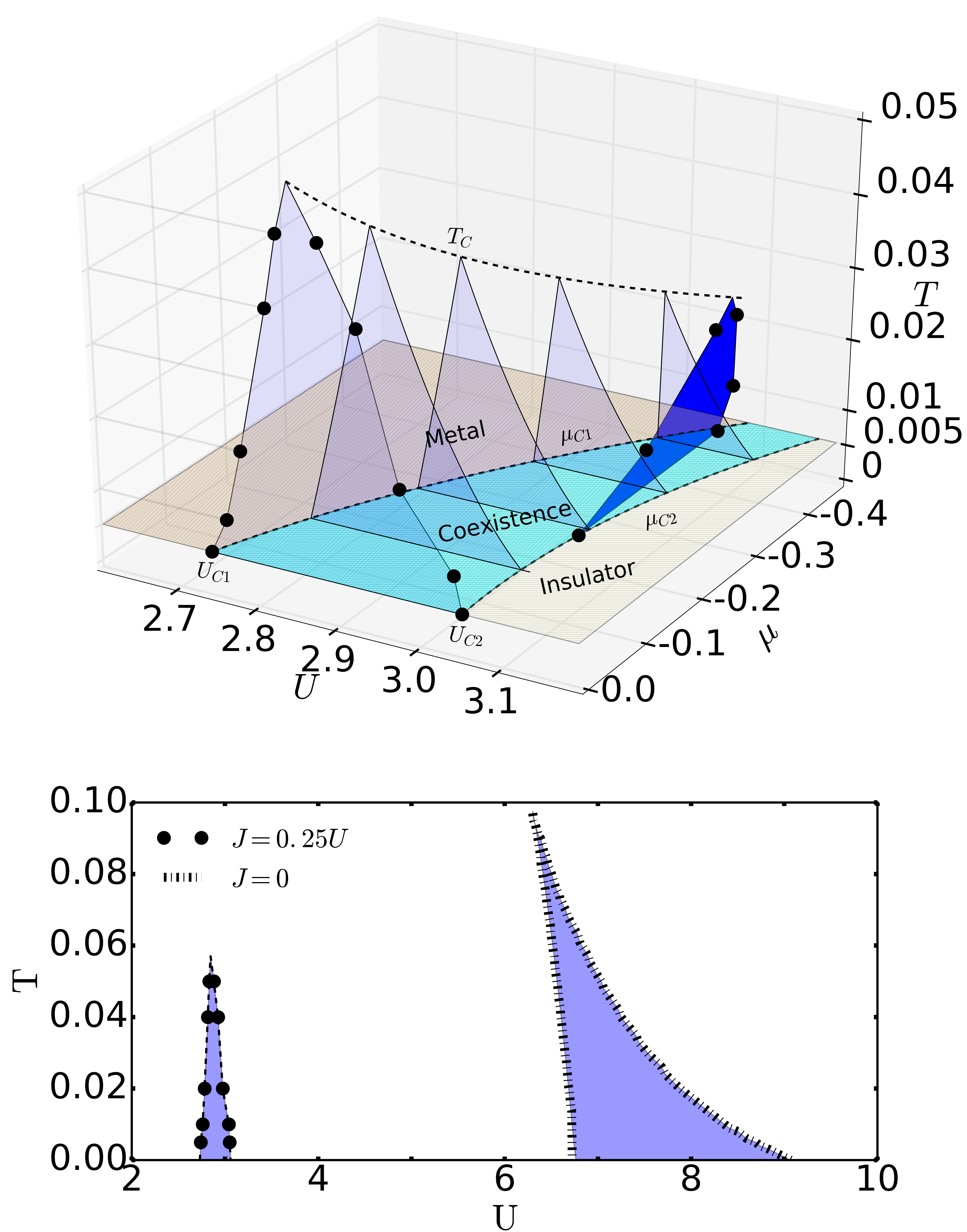}
\caption{\label{fig:2_orb_phasediag}\emph{Upper panel}: Phase diagram for the two-orbital model with density-density interaction at const. $J=0.25U$. The coexistence region constitutes a volume in the $(T,U,\mu)$ parameter-space and is indicated by the blue cuts. Black dots mark the positions of the spinoidals (defining the boundary of the coexistence region), evaluated from the calculations performed in this work. The dashed lines $\mu_{C1}$, $\mu_{C2}$ and $T_{C}$ are guides to the eye, indicating the critical chemical potentials in a plane of constant temperature ($T=0.005$ i.e. $\beta=200$, the lowest temperature we have simulated here) and the critical temperature. 
\emph{Lower panel}: Evaluated coexistence region for $J=0.25U$ (black dots) in the $(U,T)$-plane, compared to the case $J=0$ (schematic taken from Ref. \onlinecite{two_orbital_mott_transition_kawakami}).}
\end{center}
\end{figure}
\section{Results}

\subsection{\label{sec:phasedia} Phase diagram }  
Our exhaustive investigations of the two-orbital model \blue{ with density density interaction ($\alpha=0$) and $J=0.25U$} allow us to summarize our results in the form of a phase diagram, parametrized by the Hubbard interaction U, the chemical potential $\mu$ and the temperature $T$ which is presented in the upper panel of Fig. \ref{fig:2_orb_phasediag}. \blue{It may be compared to the phase diagram of the one orbital Hubbard model, which can be found in Ref. \onlinecite{Doping_1orb_wernermillis} (Fig. 1), and in Ref. \onlinecite{Jaksa_1orb} (Fig. 1); the  latter one being based on exhaustive calculations. It allows to orientate oneself in parameter space.} The cut at $\mu=0$ along the U-axis indicates the position of the coexistence region, delimited by the spinoidals  $U_{C1}(T)$ and $U_{C2}(T)$, which, upon increasing T, move closer together until they merge at $T_{c}$. \blue{Remaining at the cut $\mu = 0$, the lower panel of Fig. \ref{fig:2_orb_phasediag} compares the coexistence region, evaluated for $J=0.25U$, with a corresponding schematic for $J=0$, taken from the literature.}   The second cut \blue{in the upper panel} at constant $U=3.06$ along the $\mu$-axis shows the coexistence region delimited by $\mu_{C1}(T)$ and $\mu_{C2}(T)$; \blue{a corresponding plot for the one-orbital model can be found in the supplementary material of Ref. \onlinecite{Jaksa_1orb}}. All of the results for the two-orbital model with density-density interaction \eqref{eq:Hint} that will be presented in the following were calculated in this plane in parameter space. \par

\begin{figure}[t]
\includegraphics[width=0.45\textwidth]{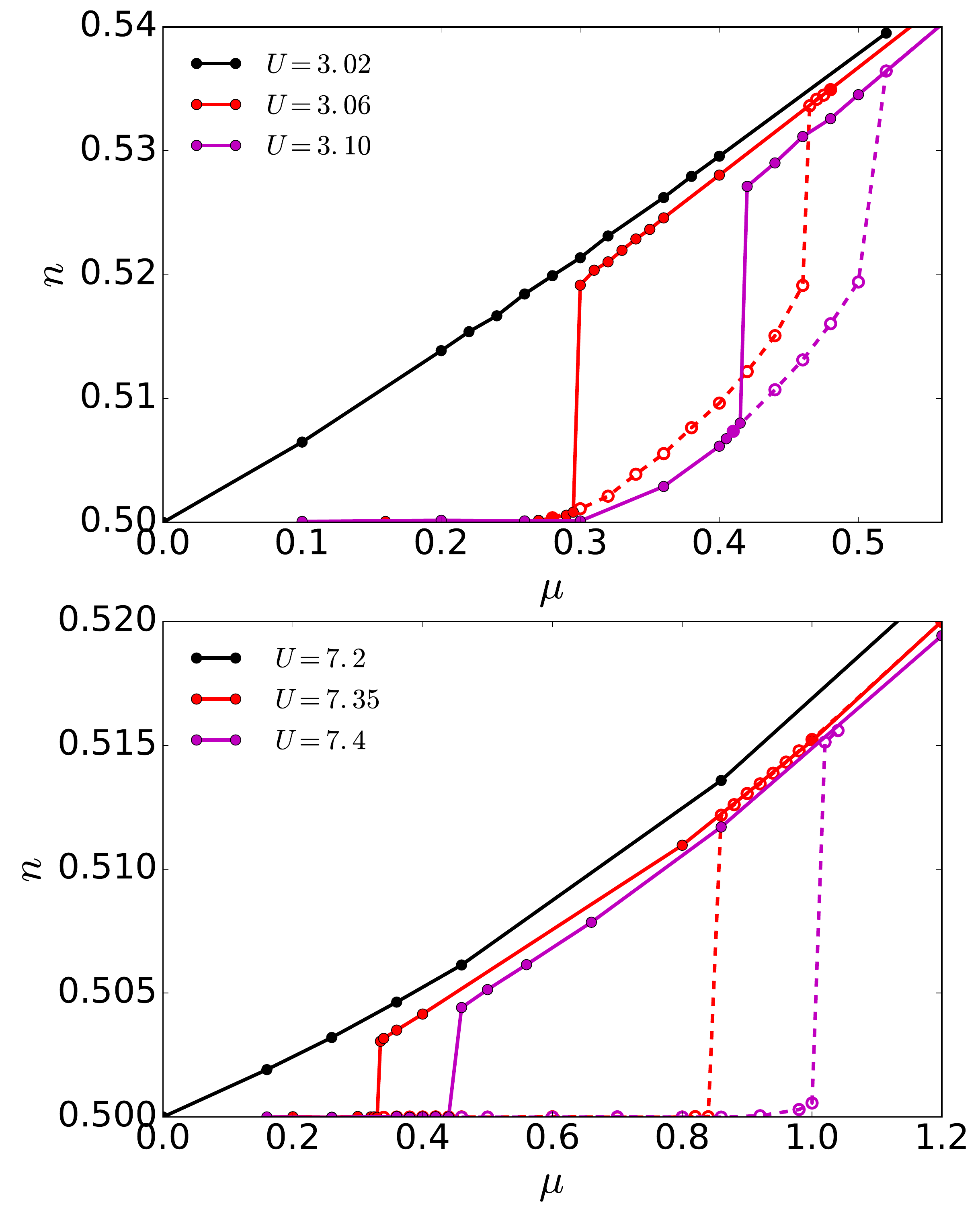}
\caption{\label{fig:2_orb_mu_nu_100} Filling $n$ per spin as a function of the chemical potential $\mu$ for the two-orbital model \eqref{eq:Hint} with  $J=0.25U$, $\beta=100$ (\emph{upper panel}) and without Hund's coupling $J=0$, $\beta=25$ (\emph{lower panel}), for various Hubbard interactions $U$. Solid(dashed) lines denote results starting from a metallic(insulating) initial configuration.}
\end{figure}

Fig. \ref{fig:2_orb_mu_nu_100} shows the filling per spin with and without Hund's coupling. \blue{Dots connected with} solid lines correspond to results obtained after starting from a metallic initial configuration, whereas \blue{dots connected with} dashed line were obtained by following the insulating branch. Compared to the case with $J=0$, the Mott transition already occurs at a strongly decreased value of $U$, which is due to the increased cost of hopping from the energetically favored configuration of aligned spins\cite{janus_faced_prl}. \blue{Also, the coexistence region can only be stabilized within a smaller range of $\mu$,} compared to the case with $J=0$ for same $\beta$. Compared to the case without Hund's coupling ($J=0$), one also has to probe much smaller temperatures to observe a first-order phase transition; an indication of a strongly reduced temperature scale $T_{K}$ (as it has also been shown in \cite{Pruschke2005} ).\par

\begin{figure}[t]
  \centering
  \hspace*{-0.3cm}
  \includegraphics[scale=0.36]{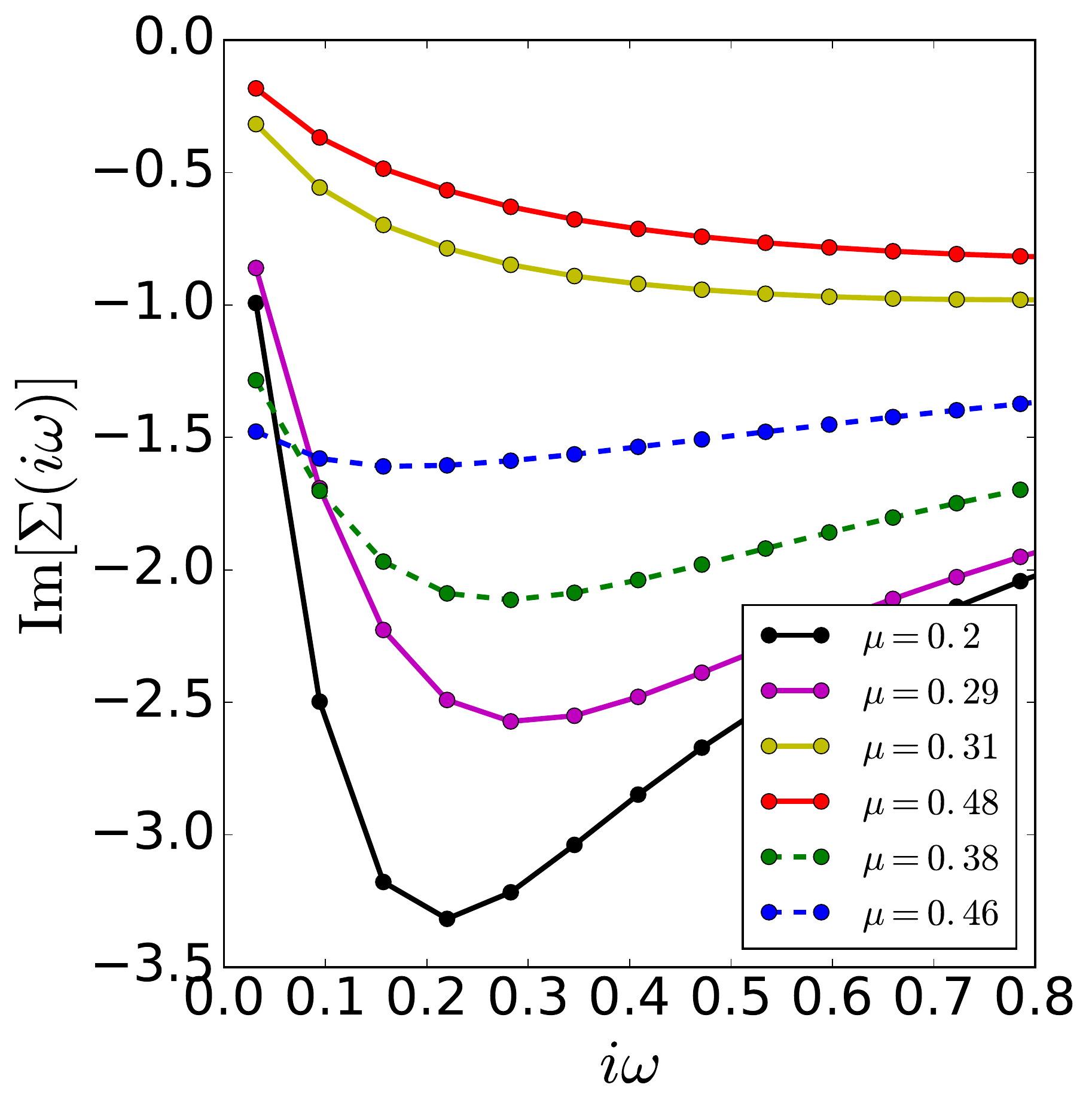}
  \includegraphics[scale=0.36]{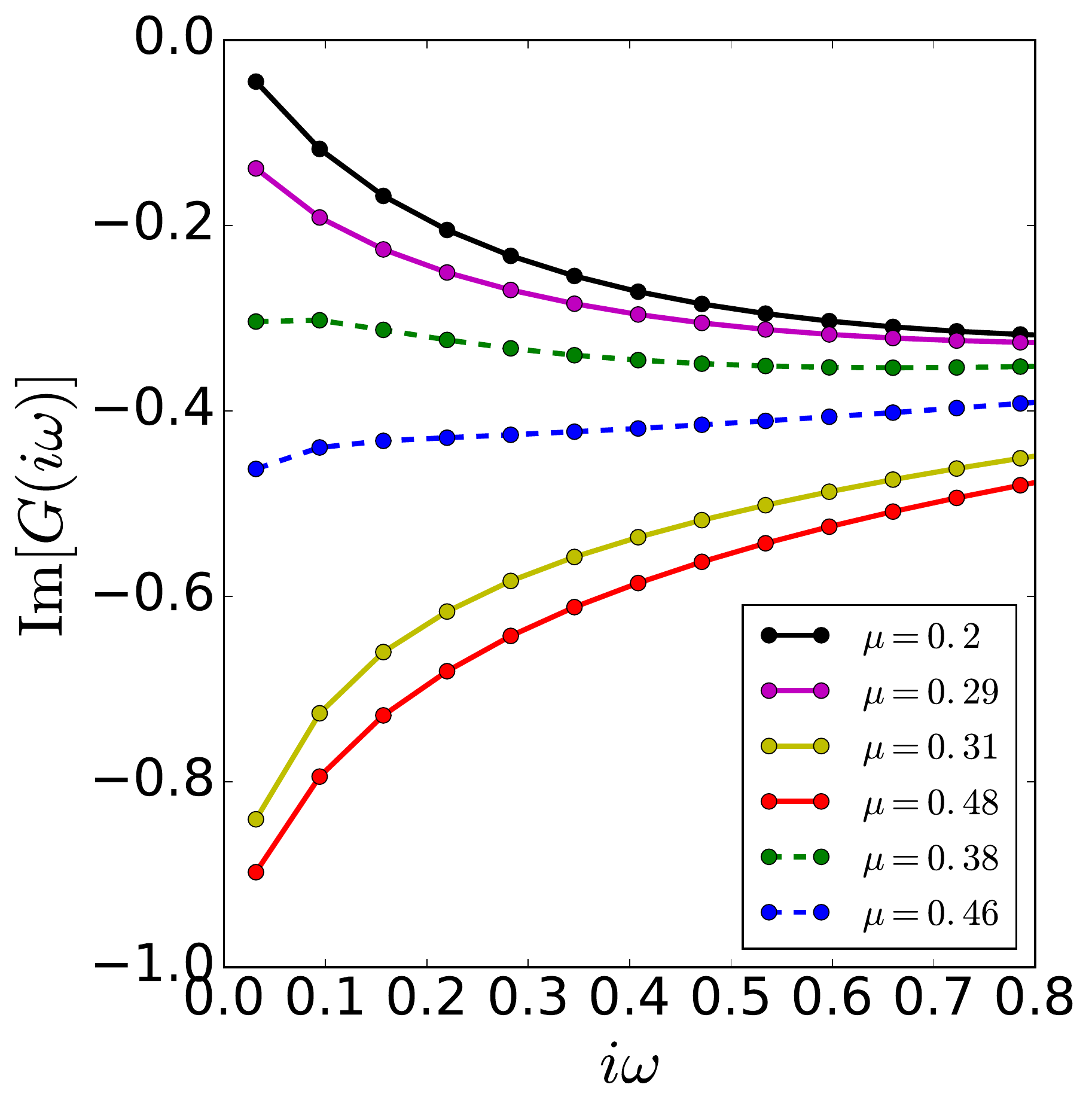}
\caption{\label{fig:sigma_G} Imaginary part of the self-energies (\emph{upper panel}) and Green's functions (\emph{lower panel}) for the two-orbital model eq. \eqref{eq:Hint} with $U=3.06$, $J=0.25U$ and $\beta=100$. Solid (dashed) lines denote simulations starting from a metallic (insulating) initial configuration.}
\end{figure}

While deviations from half filling on the insulating branch 
have already been reported\cite{compressdiv_kotliar} as due to finite temperature effects, it is remarkable how strong the system can be doped while still remaining on the insulating branch. Looking at the Fig. \ref{fig:sigma_G}, it also becomes apparent that the ``insulating'' branch is not at all insulating anymore for high dopings: While the self-energies \blue{indicate decoherence of the single-particle states }
in the coexistence region (with $\mathrm{Im}\Sigma(i\omega)$ not  \blue{ approaching} zero upon extrapolating from the lowest Matsubara frequency $i\omega_{0} = i\pi/\beta$ to zero), the imaginary part of the Green's function indicates a significant spectral weight at the Fermi level. Similar behavior has been observed in\cite{sordi_finite_doping,sordi_nature} for cluster models in two dimensions; however there, the reason for this was attributed to the effect of short-range spin correlations - considering that no similar behavior was observed in the single-site approximation\cite{compressdiv_kotliar,Doping_1orb_wernermillis}. \par

\subsection{Temperature considerations}
\begin{figure}[t]
\includegraphics[scale=0.32]{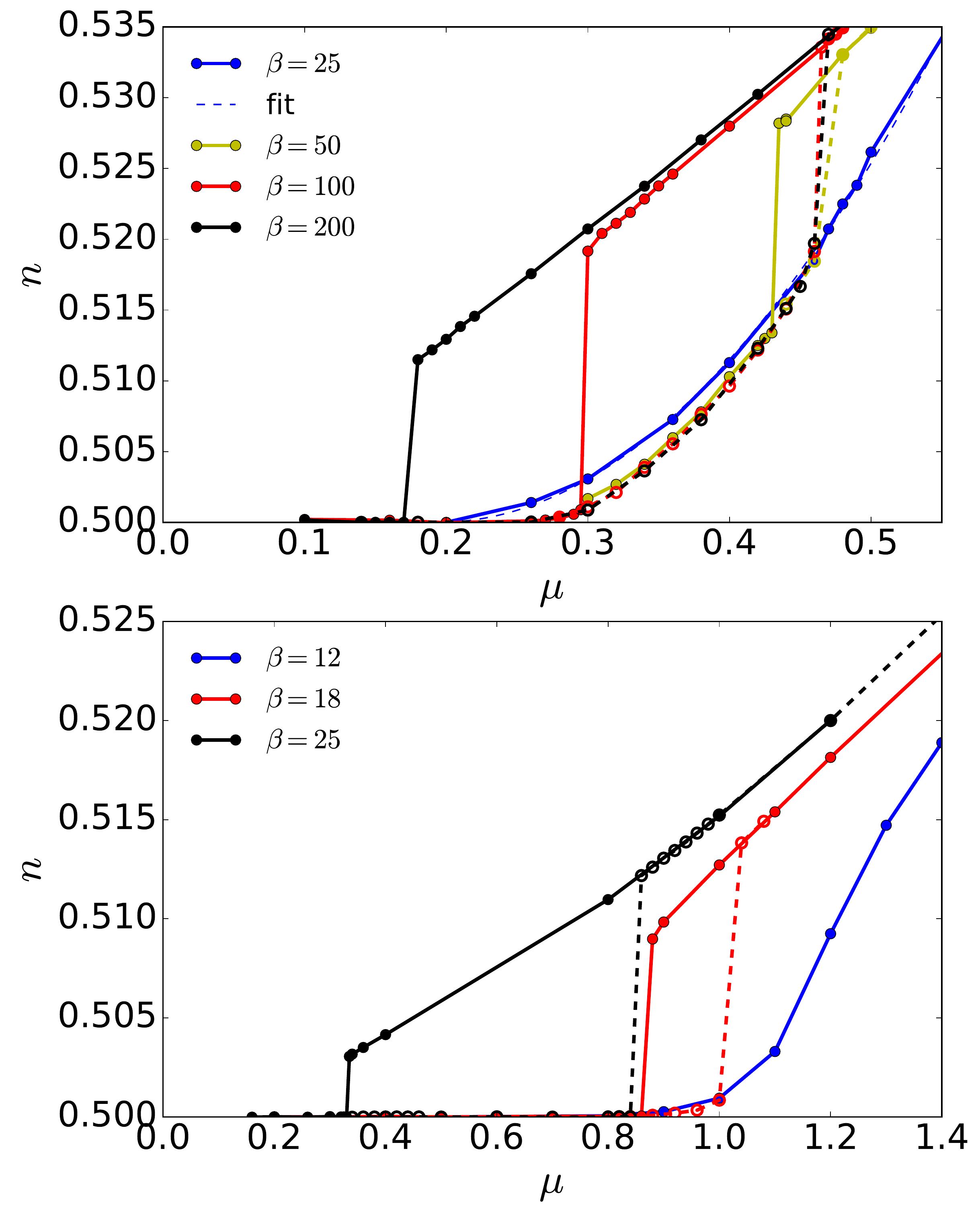}
\caption{\label{fig:2orb_mu_nu_T} \emph{Upper panel}: Filling $n$ per spin as a function of $\mu$ in the two-orbital model \eqref{eq:Hint} for different temperatures $T=1/\beta$, with $U=3.06$ and $J=0.25U$. The curve for $\beta=25$ has been fitted to $n = A(\mu-\mu_{0})^{2}$, yielding $A=0.27$, $\mu_{0}=0.19$. \emph{Lower panel}: Filling $n$ per spin as a function of $\mu$ in the two-orbital model \eqref{eq:Hint} for different temperatures $T=1/\beta$, with $U=7.35$ and $J=0$. Solid (dashed) lines designate solutions starting from a metallic (insulating) configuration.}
\end{figure}

Fig. \ref{fig:2orb_mu_nu_T} shows the electronic doping as a function of the chemical potential for $J=0.25U$ (upper panel) and $J=0$ (lower panel) for different temperatures. \blue{ Again, metallic solutions correspond to dots connected with solid lines, while insulating solutions are connected by dashed lines.}\par
At $\beta=25$ (which is already above the critical temperature for a first-order transition to occur) and finite $J=0.25U$, our line perfectly fits to $n \sim (\mu - \mu_{0})^{2}$; a behavior that has already been reported in the one-band model\cite{Doping_1orb_wernermillis}, indicating that the compressibility starts at low doping as a linear function of the chemical potential $\kappa = \partial n /\partial\mu \sim |\mu-\mu_{c}|$.

\begin{figure}[t]
\hspace*{-0.3cm}
{\includegraphics[scale=0.3]{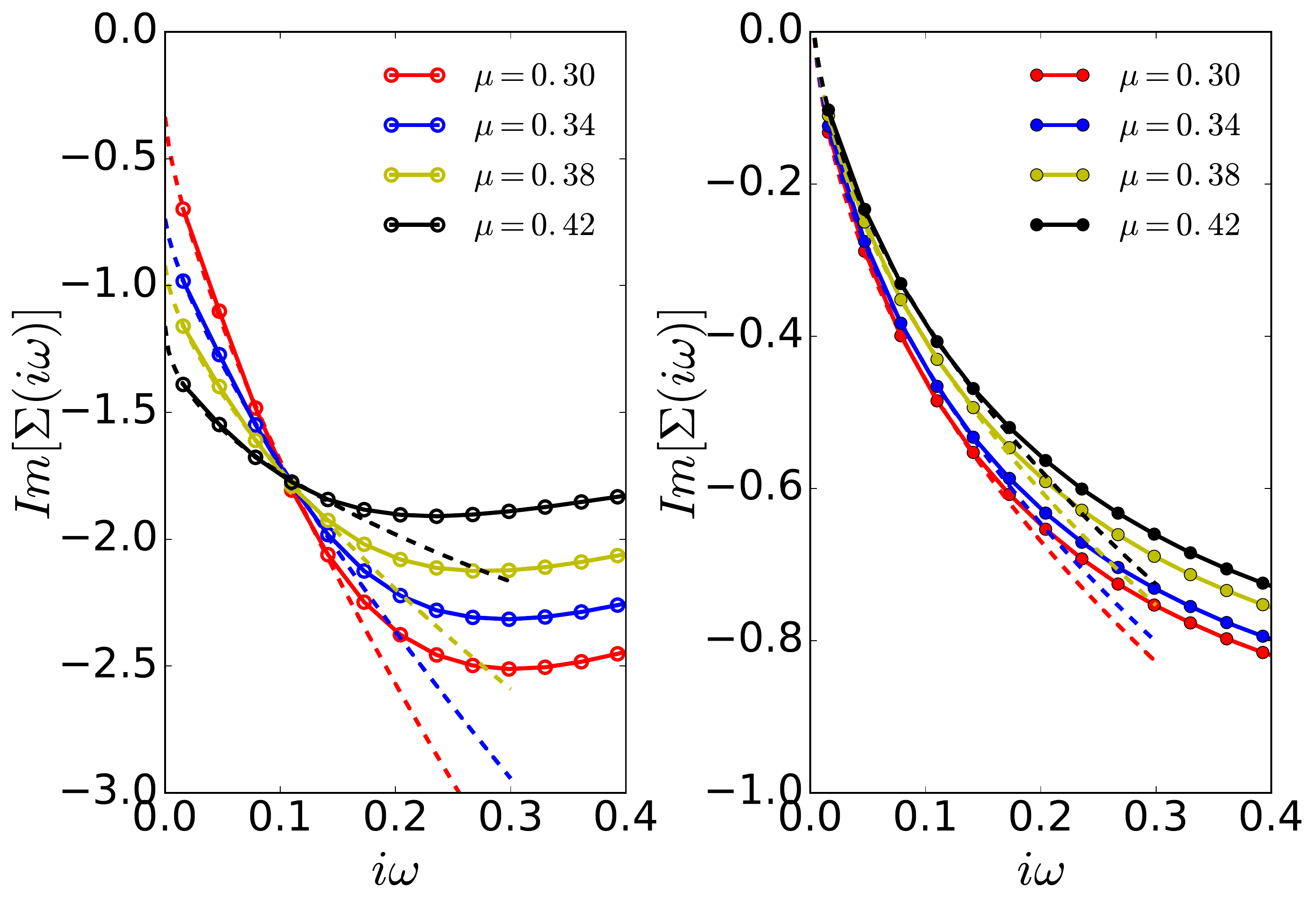}}
 \caption{\label{fig:finite_interc}%
  Self-energies of the two-orbital model for several chemical potentials $\mu$,  with $U=3.06$, $J=0.25U$ and inverse temperature $\beta=200$. Dashed lines indicate least-square fits to the function $c+b|i\omega|^{\delta}$.
  Left panel: Imaginary part of the self-energy on the Matsubara axis on the insulating branch of the coexistence region, $\delta = 0.72,0.75,0.66,0.50$ (for $\mu = 0.30,0.34,0.38,0.42$). Right panel: Same quantity on the metallic branch, $\delta = 0.45, 0.46,0.48, 0.50$ (for $\mu = 0.30,0.34,0.38,0.42$).
 }%
\end{figure}

\begin{figure}[t]
\hspace*{-0.3cm}
{\includegraphics[width=0.45\textwidth]{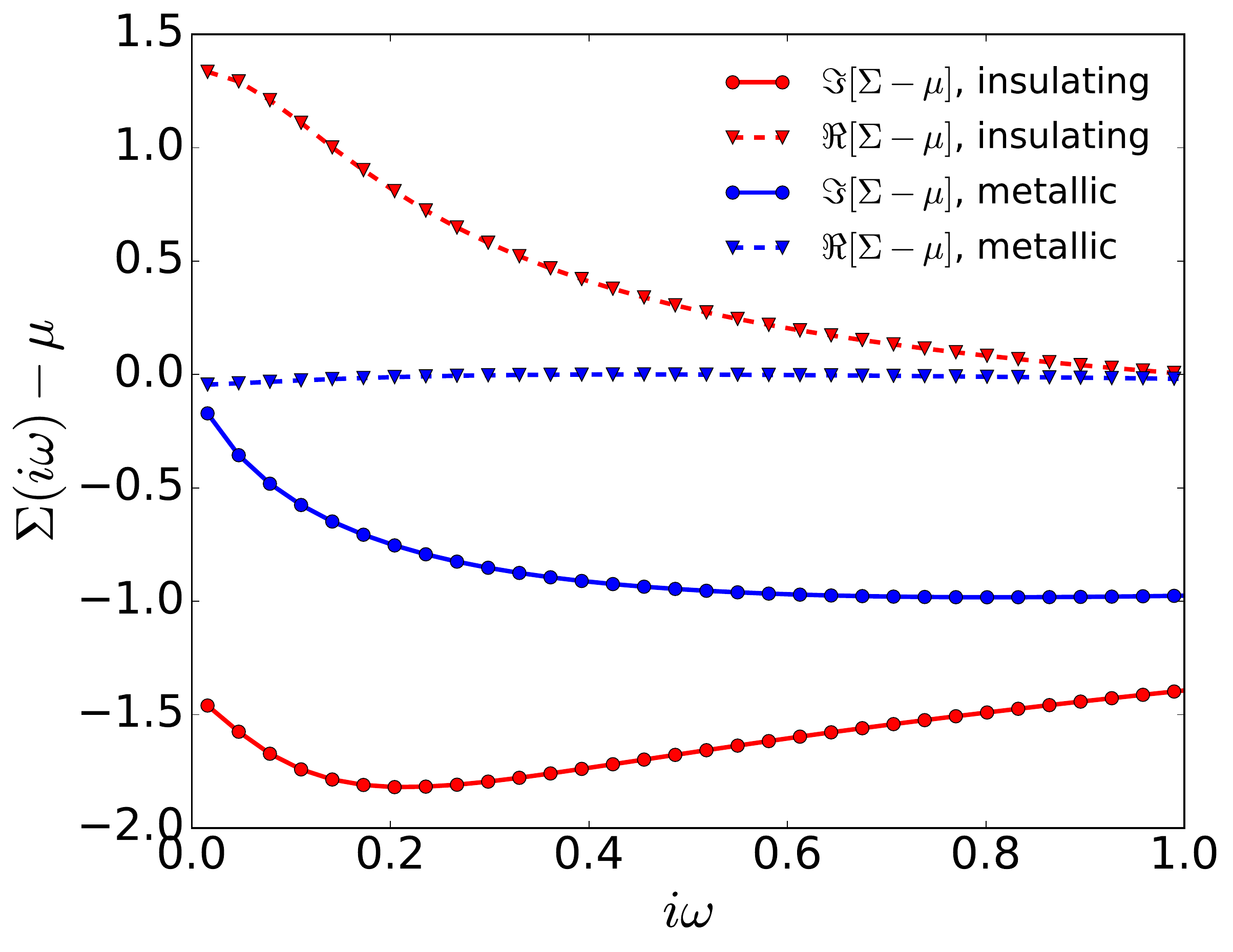}}
 \caption{\label{fig:self_energies_const_n}%
 Real and imaginary part of the self-energies on the Matsubara axis for the two-orbital model with $U=3.06$, $J=0.25U$ and $\beta=200$ for constant doping $n=0.515$/spin on the metallic (blue) and insulating (red) branch. 
 }%
\end{figure}


\begin{figure}[t]
\hspace*{-0.3cm}
{\includegraphics[width=0.45\textwidth]{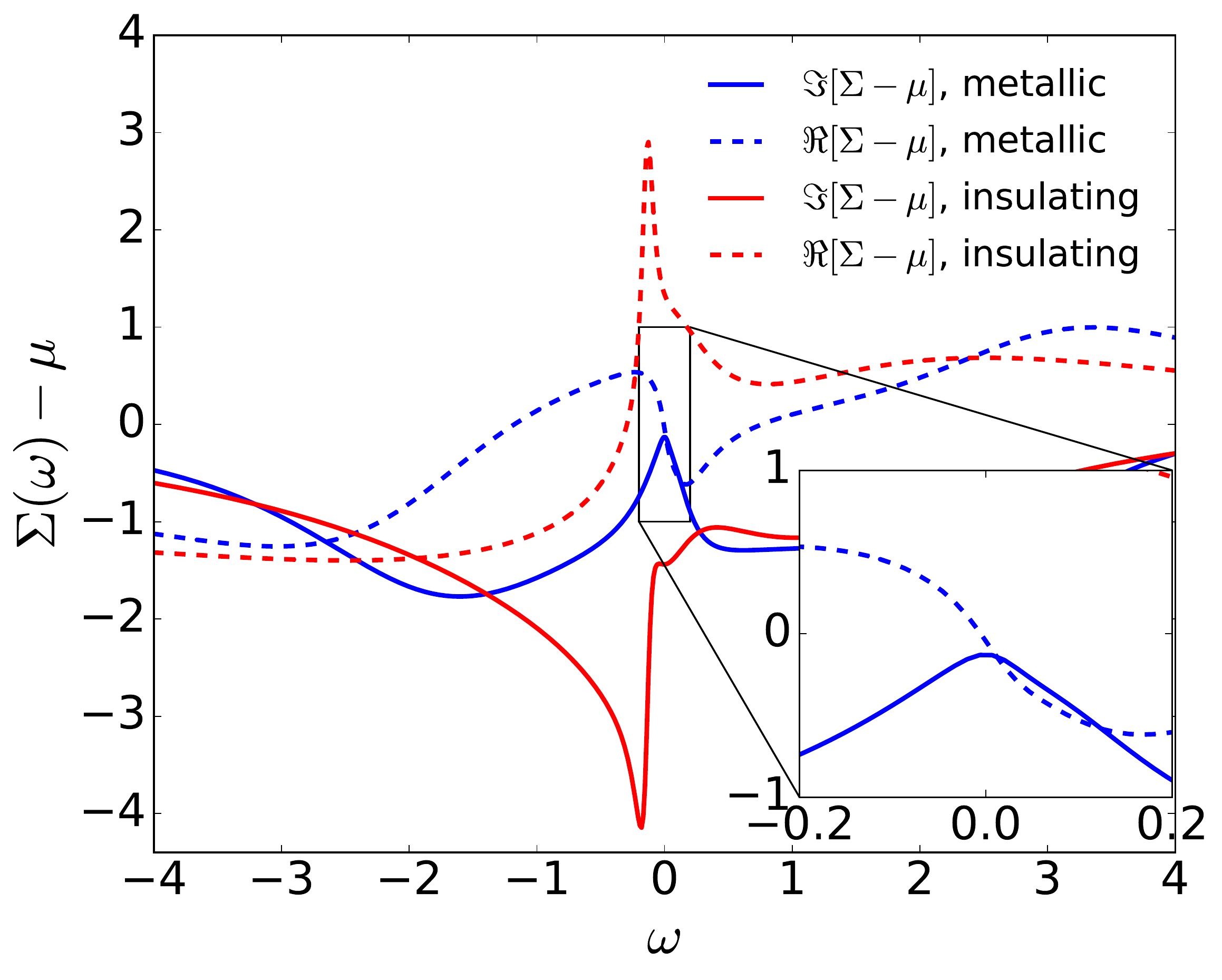}}
 \caption{\label{fig:self_energies_real}%
 \blue{ Real and imaginary part of the self-energies on the real-frequency axis for the two-orbital model with $U=3.06$, $J=0.25U$ and $\beta=200$ for doping $n=0.515$/spin on the metallic (blue) and insulating (red) branch. The inset shows a zoom to the metallic branch around the Fermi level.  }}
\end{figure}


\begin{figure}[t]
\hspace*{-0.3cm}
{\includegraphics[width=0.45\textwidth]{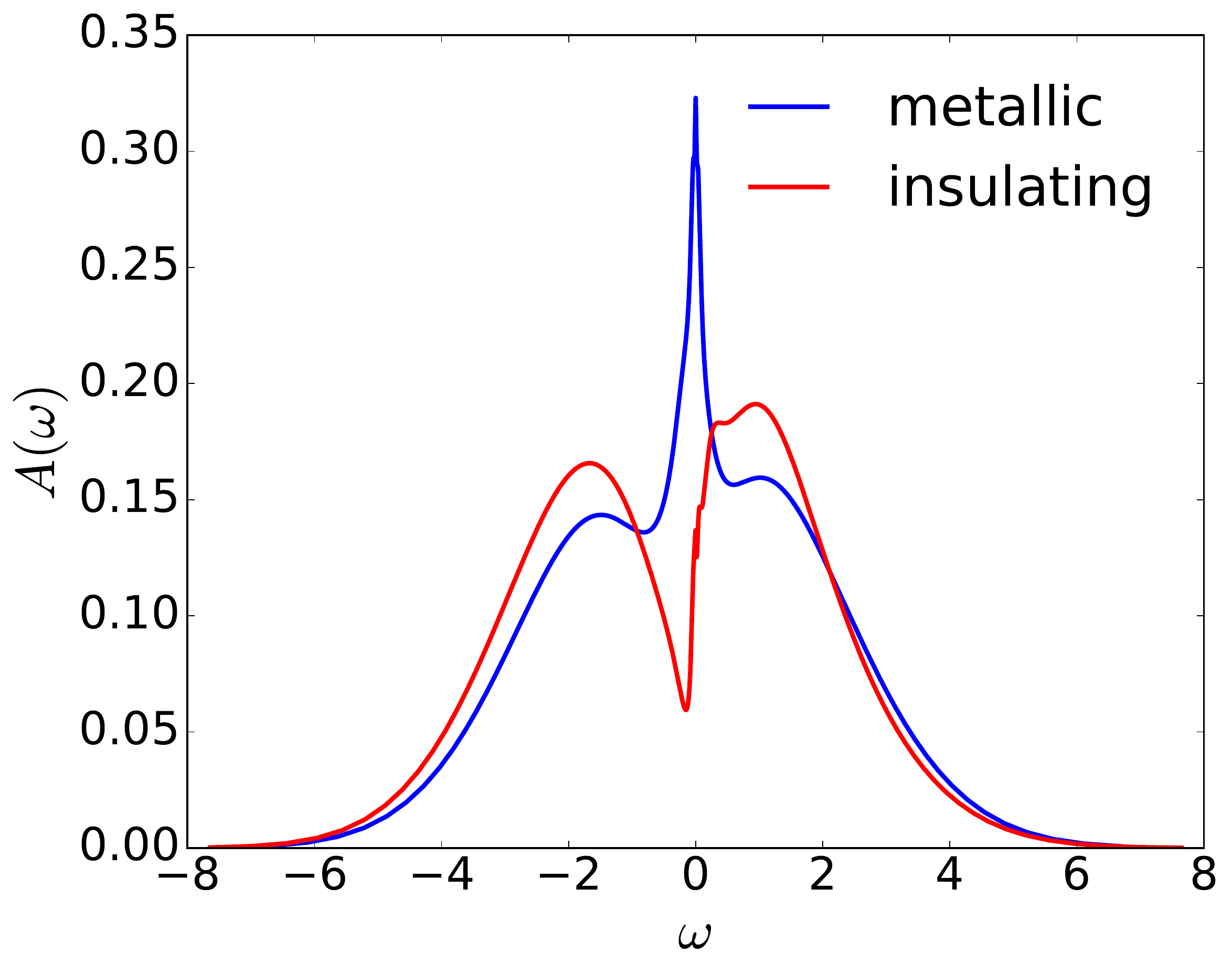}}
 \caption{\label{fig:spectral_fun}%
  Spectral functions on the metallic and insulating branch for fixed filling $n=0.515$/spin, $U=3.06$, $J=0.25$ and $\beta =200$.}
\end{figure}


\begin{figure*}[t]
{\includegraphics[scale=0.35]{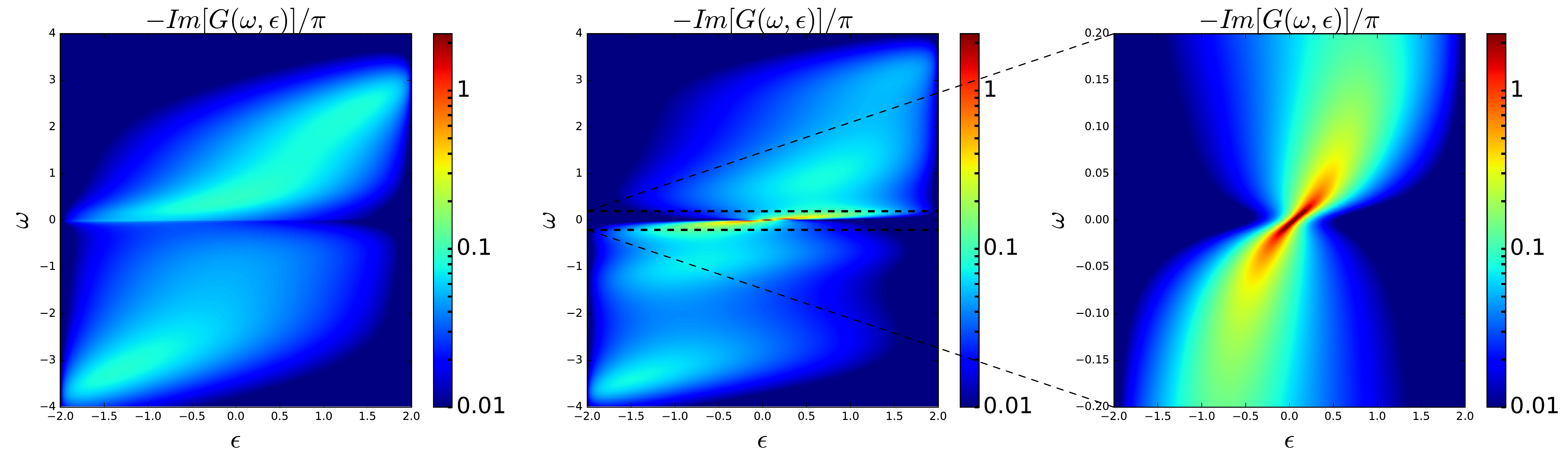}}%
 \caption{\label{fig:pseudo_dispersion}%
 Spectral functions $A(\omega,\eps) = -Im[G(\omega,\eps)]/\pi$ on the metallic and insulating branch for fixed filling $n=0.515$/spin, $U=3.06$, $J=0.25$ and $\beta =200$. The Green's functions were calculated as $G(\omega,\eps) = (\omega + \mu - \eps - \Sigma(\omega))^{-1}$ after analytic continuation of the self-energies $\Sigma(i\omega)$.
 }%
\end{figure*}

\begin{figure*}[t]
{\includegraphics[scale=0.32]{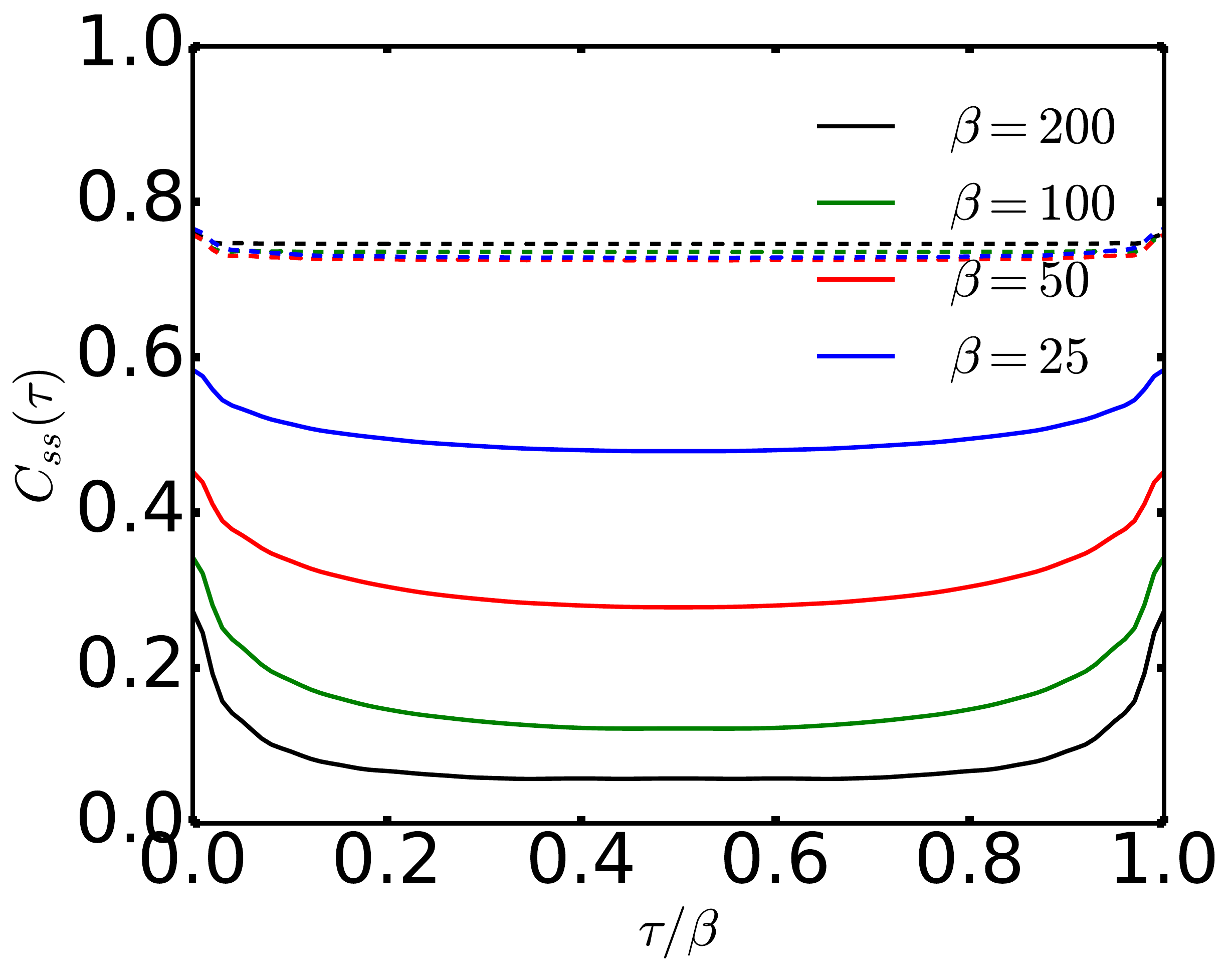}}%
{\includegraphics[scale=0.32]{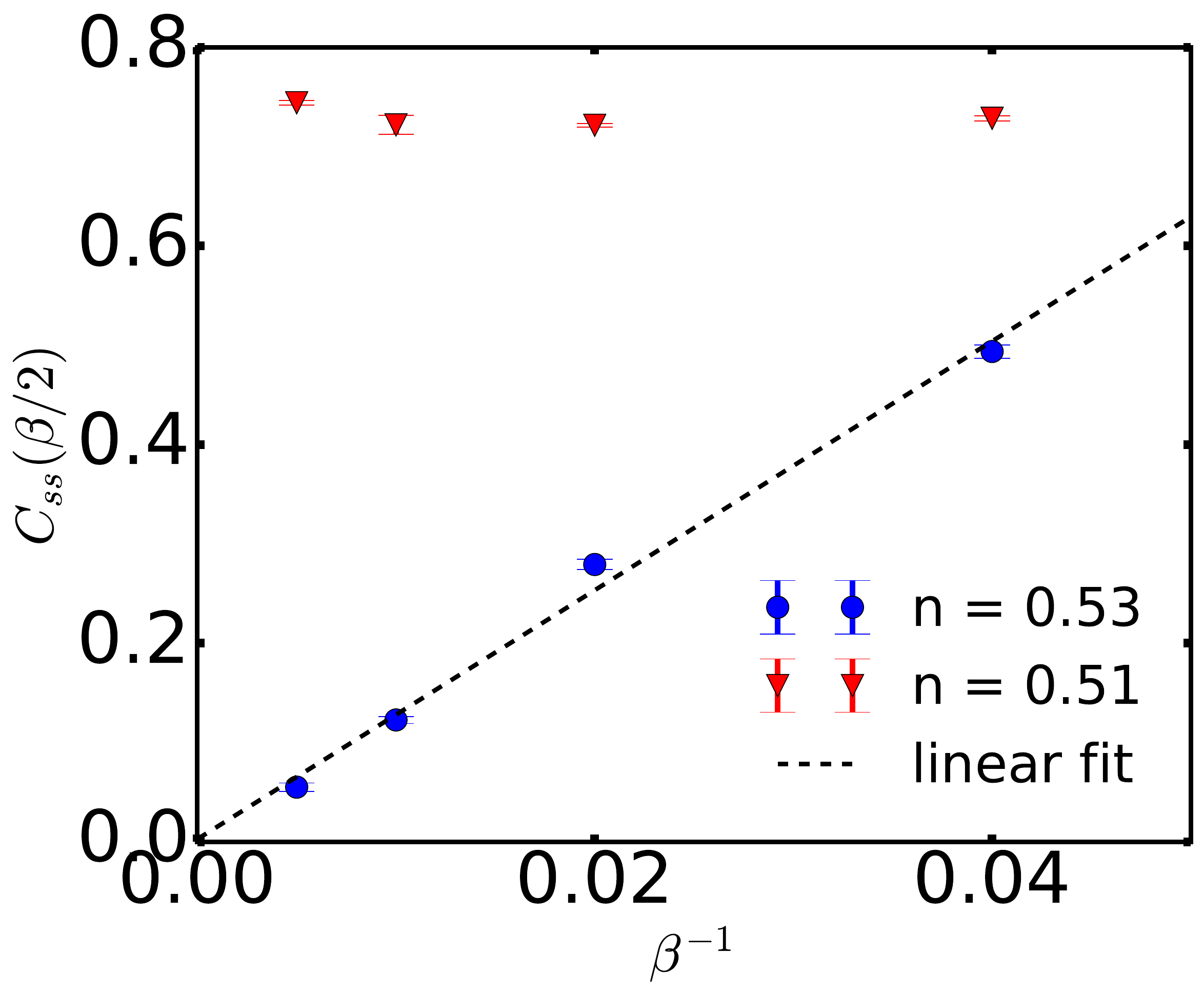}}
 \caption{\label{fig:spinspin_correl}%
  Spin-spin correlations $C_{ss}(\tau) = \Braket{S_{z}(\tau)S_{z}(0)}$ for the two-band model at $U=3.06$, with $J=0.25U$. Upper panel: The solid(dashed) lines show the spin-spin correlation functions at fillings $n=0.53$($n=0.51$) for different temperatures. Lower panel: \blue{Blue dots (red triangles)} mark the correlations at $\tau = \beta/2$ for $n=0.53$($n=0.51$) as a function of the temperature $T=\beta^{-1}$. The dashed line is a linear fit (for $n=0.53$).
 }%
\end{figure*}
\blue{ For lower temperatures, a coexistence region emerges, of which the extent in $\mu$ grows upon further lowering the temperature. In the following, we will discuss results obtained at $\beta=200$, the lowest temperature under consideration. \par
The left panel of Fig. \ref{fig:finite_interc} shows the self-energies on the branch which is adiabatically connected to the insulator. For small $|i\omega|$, we find the imaginary part of the self-energies to be characterized by a fractional power-law behavior $- \mathrm{Im}[\Sigma (i\omega)] \sim c + b|i\omega|^{\delta}$, with an exponent $\delta$ that decreases upon approaching $\mu \rightarrow \mu_{c1}$. On this branch, we also find a significantly large finite intercept $c\neq 0$, 
indicating strong decoherence. This decrease in coherence was already visible in the results shown in Fig. \ref{fig:sigma_G} for $\beta=100$. Most interestingly, decoherence gets stronger upon further doping the system, signifying enhanced scattering, while at the same time there is an increase in $|Im[G(iw_{0})]|$, testifying of a growth in spectral weight at the Fermi level. \par
The right panel of Fig. \ref{fig:finite_interc} shows the same quantities on the metallic branch. While the finite intercept of $\mathrm{Im}[\Sigma(i\omega)]$ disappeared, we still find power-law behavior with exponents $\delta \sim 0.5$.\par
The temperature corresponding to $\beta = 200$ is low enough to find a range of dopings for which both a metallic and an insulating solution can be stabilized. This allows us to perform a direct comparison of the two solutions at constant doping; the results are presented in Fig.\ref{fig:self_energies_const_n}-\ref{fig:pseudo_dispersion} for a filling of  $n=0.515$ per spin.\par
Fig. \ref{fig:self_energies_const_n} and \ref{fig:self_energies_real} compare the the self-energies of the two solutions on the Matsubara and real-frequency axis, respectively; the corresponding spectral functions
are presented in Fig. \ref{fig:spectral_fun}. 
The most distinctive feature characterizing the metallic solution in contrast to the insulating one
is a sharp resonance at the Fermi level. Such features are commonly found in the Fermi liquid regime, due to the low-energy
form of the self-energy in the Fermi liquid: $\Sigma(\omega) \sim \Sigma_{0} + (1-Z^{-1}) \omega + i(\gamma\omega^{2}  + \zeta T^{2})$ (with real $\Sigma_{0}, Z, \gamma, \zeta$). On the Matsubara axis, this
behavior translates to a linear regime of the imaginary part of the self-energy.\par
Here, the situation is different. Indeed, for the given parameters, the system is not in the Fermi liquid regime, as can be witnessed
from the self-energies in Fig. \ref{fig:finite_interc} which do not show Fermi liquid behavior. Also, the imaginary part of the self-energy corresponding to the metallic solution in Fig. \ref{fig:self_energies_const_n}
displays a fractional power-law behavior $|i\omega|^\delta$ with an exponent $\delta\approx 0.5$.
As in the case of a Fermi liquid, the imaginary parts of the self-energy approach zero up to some finite temperature corrections, leading to long-lived low-energy excitations.
Due to the non-linear behavior of $\mathrm{Im}[\Sigma(i\omega)]$ close to the Fermi level, however, no well-defined quasi-particles exist, and a formal expansion in terms of a quasi-particle
residue $Z$ would result in an energy-dependent quantity that vanishes on the Fermi surface.\par
The pronounced peak of the spectral function of the metallic solution, shown in Fig. \ref{fig:spectral_fun},
is exactly due to the vanishing imaginary part of the self-energy at $\omega=0$, as explained above. 
The insulating branch is characterized by a pseudo-gap structure, that emerges from the gapped insulator upon doping the system.  These features are further highlighted by Fig. \ref{fig:pseudo_dispersion}, which shows the spectral functions $A(\omega,\eps) = -\mathrm{Im}[G(\omega,\eps)]/\pi$, obtained as $G(\omega,\eps) = (\omega + \mu - \eps - \Sigma(\omega))^{-1}$ by analytic continuation of the self-energies $\Sigma(i\omega)$  to the real-frequency axis.
}

\subsection{Spin-spin correlations}
We now investigate the behavior of the spin-spin correlation on the two branches. To this effect, we fix the electron doping, while continuously reducing the temperature. In the case of a Fermi liquid, one expects the spin-spin correlation function 
\begin{align}
C_{ss}(\tau) = \Braket{S_{z}(\tau)S_{z}(0)}
\end{align}
to scale as $C_{ss}(\tau)\sim \frac{\beta^{-2}}{\sin (\pi \tau/\beta)^{2}}$ for low temperatures, so that its value at $\tau=\beta/2$ should behave like $C_{ss}(\beta/2) \sim \beta^{-2}$. In this case, the quantity $\beta C_{ss}(\beta/2) \sim \beta^{-1}$ is inversely proportional to the NMR relaxation time $T_1$,
and the Fermi liquid is characterized by the Korringa law $\frac{1}{T_1 T} = const$.
On the other hand, in the case of frozen spins, $C_{ss}(\beta/2)$ is expected to remain T-independent. 

Fig. \ref{fig:spinspin_correl} shows the spin-spin correlations at fixed fillings on both the metallic (solid lines) and the insulating (dashed lines) branch for various temperatures.
For $n=0.51$ we could stabilize a solution in the insulating phase for all temperatures under consideration ($T=0.04,0.02,0.01,0.005$). Despite the finite doping and the significant spectral weight at the Fermi level (see Fig. \ref{fig:sigma_G}), the correlation function at $\beta/2$ merely shows fluctuations within about the width of the error bars. Nonetheless, this might not be too surprising, considering that this branch is adiabatically connected to the Mott insulator, where one would expect such a behavior.
We also investigated the metallic branch for constant filling $n=0.53$. \blue{Here, the temperature dependence of the spin-spin correlation is almost perfectly fitted by a straight line, indicating non-Fermi liquid behavior, consistent with the comportment of the self-energies.}


The results for the 
 spin-spin correlations are particularly interesting when comparing them to the behavior of the self-energies for small Matsubara frequencies. 
In our calculations, spin-freezing was only found on the insulating branch, which is also characterized by a significant finite intercept $\mathrm{Im}\Sigma \neq 0$ when extrapolating from the lowest Matsubara frequency $i\omega_{0}=i\pi/\beta$ to zero (see Fig. \ref{fig:finite_interc}, left panel). This behavior is in accordance with the findings in Ref. \onlinecite{spinfreezing_troyermillis}, where the connection between spin-freezing and decoherence was observed for a three-orbital model. 
 

\subsection{Role of spin-flip and pair-hopping terms}
\label{non-nn}

\begin{figure}[t]
\includegraphics[width=0.45\textwidth]{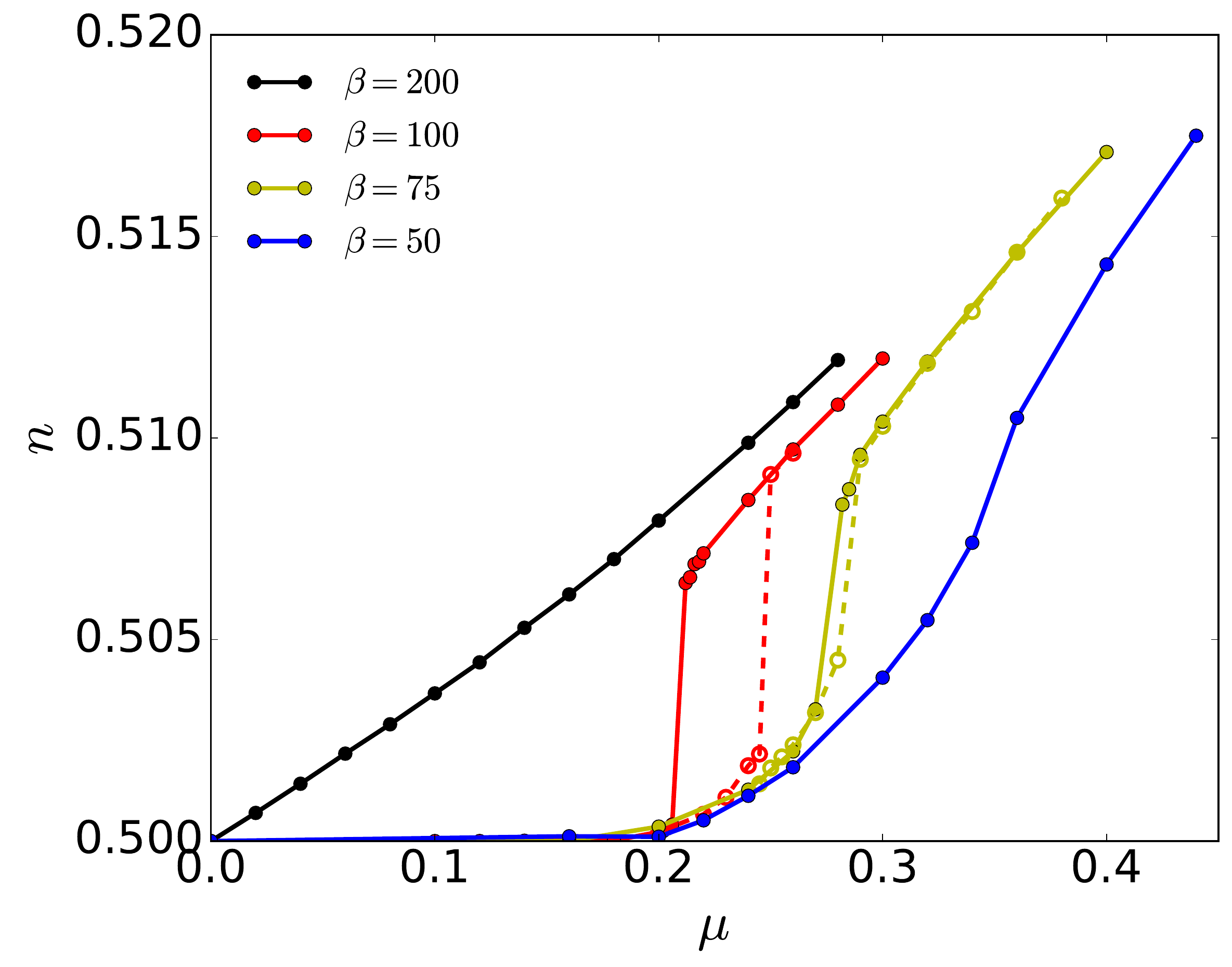}
\caption{\label{fig:kanamori}Filling $n$ per spin as a function of the chemical potential $\mu$ for the two-orbital model  with spin-flip and pair-hopping terms ($\alpha = 1$) for various temperatures, $U=3.4$ and $J=0.25U$. Solid (dashed) lines denote results starting from a metallic (insulating) initial configuration.}
\end{figure}

\begin{figure}[t]
\includegraphics[width=0.45\textwidth]{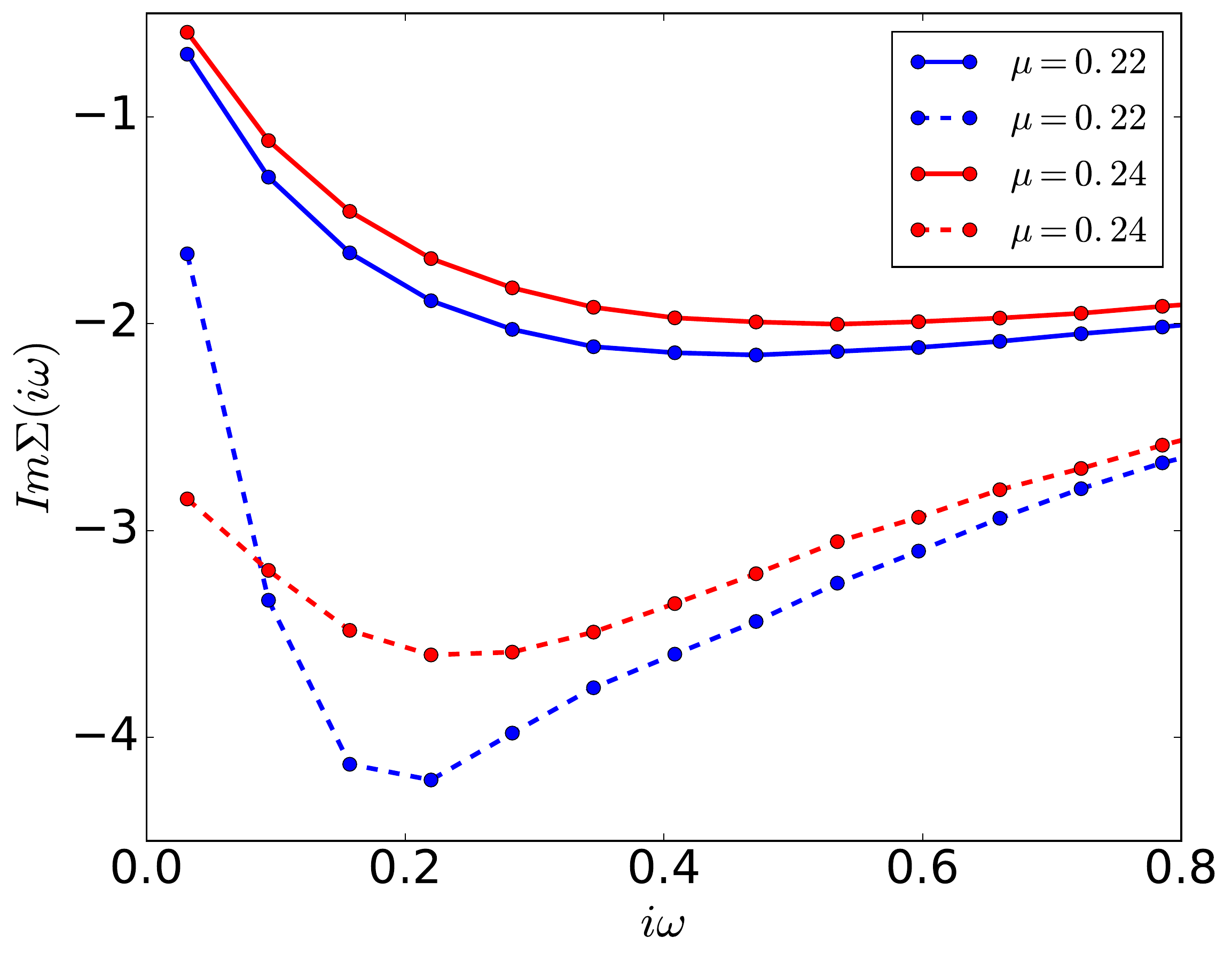}
\caption{\label{fig:kanamori_selfenergy}Imaginary part of the self-energies for the two-orbital model with spin-flip
      and pair-hopping terms in the interaction Hamiltonian, i.e. $\alpha = 1$.
  $U=3.4$, $J=0.25U$ and $\beta=100$; solid (dashed) lines denote results starting from a metallic (insulating) initial configuration.}
\end{figure}

We now turn to an investigation of the influence of the spin-flip and
pair-hopping terms, setting $\alpha=1$ in Hamiltonian (\ref{eq:Hint}).
Fig. \ref{fig:kanamori} shows
results for different temperatures.
As before, the curves representing the filling as a function of the
chemical potential split into two branches, \blue{ corresponding to the insulating and metallic solution, respectively. }
The new terms induce additional fluctuations and therefore enhance the critical interaction $U_{c2}(T)$, while the region where only an insulating solution is found is narrowed. As in the case with density-density interactions ($\alpha=0$), the insulating branch still extends to considerable dopings, although the effect is less pronounced than in the case with density-density interactions. \blue{Most strikingly, the apparent stability of the insulating branch upon changes of temperature, as witnessed in the upper panel of Fig. \ref{fig:2orb_mu_nu_T}, is not reproduced in the model with rotationally symmetric interactions.} As with the density-density Hamiltonian, 
the corresponding self-energies are characterized by non-Fermi liquid power law behavior on both branches and finite intercepts on the  insulating branches, as can be seen in Fig. \ref{fig:kanamori_selfenergy} . 

\section{Conclusion and Outlook}
In this work, we have  studied the finite temperature properties of two-orbital Hubbard models with Hund's density-density interactions, which we complemented with calculations for the rotationally symmetric case (results for the three-orbital model with density-density interaction are discussed in Appendix \ref{sec:three_orb}). As in the one-orbital case, we found a coexistence of two solutions near the Mott transition; one connected to the Fermi liquid and the other one connected to the Mott insulator. We found that at finite Hund's coupling $J$, the insulating branch can be stabilized up to high dopings, where it is characterized by \blue{ almost temperature independent} spin-spin correlations, power-law behavior of the self-energy  with increasing decoherence and a significant spectral weight at the Fermi level.\\
\blue{ For sufficiently low temperatures, we found a parameter regime in which two coexisting solutions could be stabilized at the same doping, which is particularly interesting considering possible implications for the Fermi/Luttinger surface of such systems.} For $T=0$, Luttinger's theorem \cite{Luttinger_Ward_1960} dictates the volume defined by $\mathrm{Re} G(\omega=0,\mathbf{k})>0$ 
to be proportional to the number of electrons $n$. Within methods beyond DMFT, able to access the momentum dependence of the self-energy, the two different solutions should, therefore, be distinguished by different
 shapes of the Luttinger surfaces, defined by $\mathrm{Re} G(\omega=0,\mathbf{k})=0$, while keeping its enclosed volume constant. Even if the remarkable extension of the insulating branch to finite dopings  cannot \blue{be} expected to survive the limit of $T\rightarrow 0$ \footnote{The ostensible stability upon decreasing $T$ might only be due to the exponentially slow temperature scaling, consistent with the literature\cite{deLeoPhD, Hundreview, Hund_stadler, STADLER2018}}, \blue{these findings remain interesting, especially considering realistic systems.}
\blue{ Clearly,} the low-temperature behavior of the insulating branch constitutes an interesting subject of further investigations.\par
We found that the spin-freezing regime appears to be connected to the insulating branch, while on the metallic branch spin-spin correlations decay upon decreasing the temperature. \blue{This is particularly interesting since it would imply a first-order transition between the two regimes, close to the critical interaction at half filling. Here, future investigations might focus on the nature of spin-spin fluctuations close to the endpoint $T_{c}$ of the first order transition.}\par
Earlier works\cite{spinfreezing_troyermillis} on three-orbital models found the spin-freezing phenomena to persist even in regions far away from half filling (for $U$ significantly bigger than $U_{C2}$). To explain how these two observations fit together, one must consider that 1) our main focus was on the study of two-orbital models, while for three-orbitals (Appendix \ref{sec:three_orb}, with density-density interactions), the insulating branch was found to extend to even higher dopings and that 2) our calculations were performed at $T<T_{c}$, which is not necessarily the case for all values of $U,\mu$ under consideration  in Ref. \onlinecite{spinfreezing_troyermillis}, since $T_{c}(U)$ decreases upon increasing $U$.\par
While our calculations for rotationally symmetric interactions (Fig. \ref{fig:kanamori}) seem to indicate a divergence of the compressibility upon approaching $T_{c}$, corresponding to the case with $J=0$\cite{compressdiv_kotliar}, such a tendency is not apparent in the case of density-density interactions (see Fig. \ref{fig:2orb_mu_nu_T}, upper panel). Since the theoretical arguments presented by Kotliar et al.\cite{compressdiv_kotliar} should apply independently of the specific type of the interaction, future investigations might focus on a careful analysis of the compressibility in the vicinity of $T_{c}$.\par
In a recently conducted slave-spin study \cite{Slavespin_medici} a compressibility enhancement and divergence on the metallic branch at low doping due to Hund's coupling was found at T=0 for $U>U_{c2}(T=0)$, indicating a tendency of the system towards phase separation in that regime. This effect is not observed in the present study, which however reports mainly results for $U<U_{c2}(T=0)$ (to the extent we could exactly locate $U_{c2}(T=0)$, which is very computationally expensive with the present finite-T method). In selected cases, however, as e.g. that of Fig. \ref{fig:2_orb_mu_nu_100}, upper panel, we observed a range of densities that could not be stabilized for any of the two branches. This indicates indeed a phase separation, and might be the finite-temperature signature of the physics reported in Ref. \onlinecite{Slavespin_medici}. Therefore, chances remain that a diverging compressibility might become visible at lower temperatures, especially if one considers the strong reduction of the effective Fermi-liquid temperature scale upon finite Hund's coupling. 
Future works might investigate on this effect, probing different regimes ($U>U_{c2}(T=0)$), potentially using different methods.\par
\blue{ This finally leads us to a discussion of the computational methods. The CT-Hyb quantum Monte Carlo algorithm used for our calculations is a versatile finite temperature quantum impurity solver, applicable for arbitrary interaction Hamiltonians. Despite its successes, however, some drawbacks exist: 1) The computational time scales as $\sim\beta^3$, rendering low temperature calculations increasingly costly and 2) the solver is based on an imaginary time formalism, necessitating analytic continuation procedures to retain spectral quantities. It has been shown\cite{Hund_stadler}, that these two problems can be circumvented by considering the numerical renormalization group (NRG) solver, which allows accessing arbitrarily low temperatures without much additional computational effort, directly returning real-frequency quantities. Considering future investigations, this solver is particularly interesting for models with high symmetries -- as in a recent study on SU(N) Hubbard models\cite{,Multi_orbital_von_Delft} -- since those can be exploited to improve efficiency. 
}

\section{Acknoledgments}
We thank Michel Ferrero for help with the TRIQS toolbox\cite{TRIQS}, Yusuke Nomura for providing his quantum Monte Carlo code, as well as Jernej Mravlje and Masatoshi Imada for fruitful
discussions. This work was supported by the European Research Council (Consolidator Grant No. 617196 CORRELMAT and Consolidator Grant No. 724177 StrongCoPhy4Energy) and supercomputing time 
at IDRIS/GENCI Orsay (Project No. t2018091393).
We thank the computer team at CPHT for support.

\appendix 
\section{Three-orbital case}\label{sec:three_orb}
\begin{figure}[t]
\includegraphics[width=0.45\textwidth]{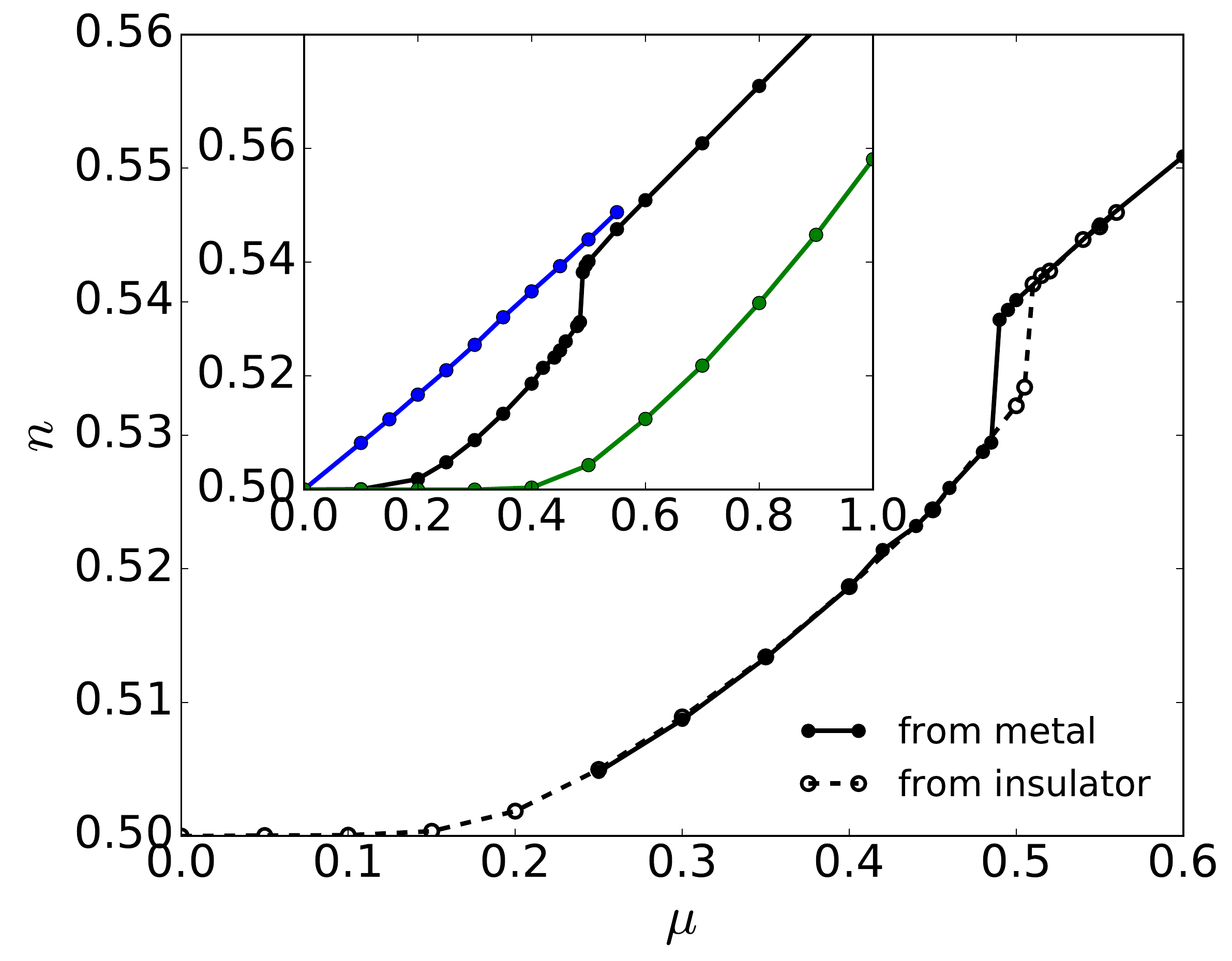}
\caption{\label{fig:3orb_mu_nu_100}Filling $n$ per spin as a function of the chemical potential $\mu$ for a three-band model with $U=2.3$, $J=0.25U$ and $\beta=50$. Solid (dashed) lines denote results starting from a metallic (insulating) initial configuration. Inset: $n(\mu)$ per spin for $U=2.25,2.3,2.6$ (blue, black, green) for $J=0.25$, starting from a metallic configuration.}
\end{figure}

Fig. \ref{fig:3orb_mu_nu_100} shows the equivalent to Fig. \ref{fig:2_orb_mu_nu_100} for the three-band model. In contrast to models without Hund's coupling (where  $T_{K}/D \sim (T_{K}^{N=1}/D)^{1/N}$, with $D=2t$ being the half bandwidth), $T_{K}$ is rather decreased for more orbitals in the case of $J>0$. However, the increased cost of performing numerical simulations yet prevented us from exploring very low temperatures. While the coexistence region is smaller than in the two-band case, the metal-insulator transition is characterized by an insulating branch that can be stabilized to even higher dopings. In general, the phenomenology seems to be the same, more extensive research would, however, be of interest.

\nocite{*}

%

\end{document}